\documentstyle[12pt,osa]{revtex}
\parskip=\medskipamount

\title{On possible generalization of the superstring action to eleven
dimensions}
\author{A.A. Deriglazov\thanks{deriglaz@fma.if.usp.br}
and A.V. Galajinsky\thanks{galajin@fma.if.usp.br}\\}
\address{Instituto de F\'\i sica, Universidade de S\~ao Paulo,\\
P.O. Box 66318, 05315-970, S\~ao Paulo, SP, Brasil}
\begin{document}
\maketitle

\begin{abstract}
We suggest a $D=11$ super Poincar\'e invariant action for the
superstring which has free dynamics in the physical variables sector.
Instead of the standard approach based on the searching for an action
with local $\kappa$-symmetry (or, equivalently, with corresponding
first class constraints), we propose a theory with fermionic
constraints of second class only. Then the $\kappa$-symmetry and
the well known $\Gamma$-matrix identities are not necessary for the
construction. Thus, at the classical level, the superstring action of the
type described can exist in any spacetime dimensions and the known
brane-scan can be revisited.
\end{abstract}

\noindent
{\bf PAC codes:} 0460D; 1130C; 1125\\
{\bf Keywords:} eleven dimensional superstring, covariant quantization.

\section{Introduction}

A revival of interest to the problem of covariant formulation for eleven
dimensional superstring is due to the search for M-theory (see
Refs. 1--5 and references therein) which is expected to be the underlying
quantum theory for the known extended objects. In the strong coupling limit
of M-theory $R^{11}\to\infty$, where $R^{11}$ is the radius of
11th dimension, the vacuum is eleven dimensional Minkowski and the
effective field theory is $D=11$ supergravity. Up to date, $D=11$
supergravity is viewed as the strong coupling limit of the ten dimensional
type IIA superstring \cite{1}.  Since $D=11$ super Poincar\'e symmetry
survives in this special point in the moduli space of M--theory
vacua ("uncompactified M--theory" according to Ref. 5), one may
ask of the existence of a consistent $D11$ quantum theory with $D=11$
supergravity being its low energy limit. One possibility is the
supermembrane action \cite{6,7,8}, but in this case one faces with the
problem of a continuous spectrum for the first quantized supermembrane
\cite{9,10}. By analogy with the ten dimensional case, where the known
supersymmetric field theories can be obtained as the low energy limit
of the corresponding superstrings \cite{5}, the other natural candidate
might be a $D=11$ superstring theory. But the problem is that a
covariant formulation for $D=11$ superstring action is unknown even at
the classical level. The classical Green--Schwarz (GS) superstring
(with manifest space-time supersymmetry and local
$\kappa$-symmetry) can propagate in 3, 4, 6 and 10 spacetime dimensions
\cite{11} and the standard approach fails to construct a $D=11$
superstring action.

The crucial ingredient in the construction of the GS superstring action
is the $\Gamma$-matrix identity
\begin{equation}
\Gamma^\mu_{\alpha(\beta}(C\Gamma^\mu)_{\gamma\delta)}=0.
\end{equation}
It provides the existence of both global supersymmetry and local
$\kappa$-symmetry for the action \cite{11,12}. The
$\kappa$-symmetry, in its turn, eliminates half of the initial
$\theta$--variables as well as provides free dynamics in the physical
variables sector.  In this paper we discuss a possibility to construct
a classical superstring action with those two properties in eleven
dimensions. Subsequent development of our method may shed light on the
problem of constructing the corresponding quantum theory. To elucidate
the construction which will be suggested below let us discuss the
problem in the Hamiltonian framework, where one finds the well known
fermionic constraints $L_\alpha=0$ (see, for example, Refs. 11 and 12)
which obey the Poisson brackets
\begin{equation}
\{ L_\alpha,L_\beta\}=2i(\hat p^\mu+\Pi^\mu_1)\Gamma^\mu_{\alpha\beta}
\delta(\sigma-\sigma')-2\bar\theta^\gamma\partial_1\theta^\delta
\Gamma^\mu_{\gamma(\delta}(C\Gamma^\mu)_{\alpha\beta)}
\delta(\sigma-\sigma').
\end{equation}
By virtue of Eq. (1), the last term in Eq. (2) vanishes for
$D=3,4,6,10$. The resulting equation then means that half of the
constraints are first class, which exactly corresponds to the
$\kappa$-symmetry presented in the Lagrangian framework.

The next step is to impose an appropriate gauge. Then the following
set of functions:
\begin{eqnarray}
&& L_\alpha=0,\\
&& \Gamma^+\theta=0,
\end{eqnarray}
is a system of second class (even through Eq. (1) has not been used).

The situation changes drastically for the $D=11$ case, where instead of
Eq. (1) one finds \cite{13,14,15}
\begin{equation}
10 \Gamma^\mu_{\alpha(\beta}(C\Gamma^\mu)_{\gamma\delta)}+
\Gamma^{\mu\nu}_{\alpha(\beta}(C\Gamma^{\mu\nu})_{\gamma\delta)}=0.
\end{equation}
Being appropriate for the construction of the supermembrane action [6],
this identity does not allow one to formulate a $D=11$ superstring with
desirable properties. As was shown by Curtright \cite{13}, the globally
supersymmetric action based on this identity involves additional to
$x^i$, $\theta_a$, $\bar\theta_{\dot a}$ degrees of freedom in the
physical sector. Moreover, it does not possess a $\kappa$-symmetry
that could provide free dynamics \cite{13,14}.

In this paper we suggest a $D=11$ super Poincar\'e invariant action
for the classical superstring which has free dynamics in the
physical variables sector. Instead of the standard approach based on
the searching for an action with local $\kappa$-symmetry (or,
equivalently, with corresponding first class constraints), we
present a theory in which covariant constraints like Eqs. (3), (4)
arise among others. Since it is a system of second class constraints,
$\kappa$-symmetry and the identity (5) are not necessary for the
construction. Thus, at the classical level, a superstring of the type
described can exist in any spacetime dimension and the known brane scan
\cite{4} can be revisited. For definiteness, in this paper we discuss
the $D=11$ case only.

Two comments are in order. First, one needs to covariantize Eq. (4).
The simplest possibility is to introduce an auxiliary variable
$\Lambda^\mu(\tau,\sigma)$ subject to $\Lambda^2=0$ and replace Eq. (4)
by $\Lambda_\mu\Gamma^\mu\theta=0$. The most preferable formulation
seems to be that in which the gauge $\Lambda^-=1$ is possible.
Then Eq. (4) is reproduced. Unfortunately, it seem to be impossible
to introduce a pure gauge variable with the desired properties
\cite{16,17,18,19,20}. Below, we present a formulation in which only
zero modes of auxiliary variables survive in the sector of physical
degrees of freedom.  Since states spectrum of a string is determined by
the action on the vacuum of oscillator modes only, one can expect that
the presence of the zero modes will be inessential for the case. This
fact will be demonstrated within the canonical quantization framework
in Sec. 2 and 4.

Second, one expects that a model with constraints like Eqs. (3), (4)
will possess (if any) off-shell super Poincar\'e symmetry in a
nonstandard realization. Actually, global supersymmetry which does not
spoil the equation $\Lambda_\mu\Gamma^\mu\theta=0$ looks like
$\delta\theta\sim\Lambda_\mu\Gamma^\mu\epsilon$. On-shell, where
$\Lambda^2=0$, only half of the supersymmetry parameters
$\epsilon^\alpha$ are essential.

It is worth mentioning another motivation for this work. As was shown
in Refs. 21--25, an action for the super $D$-brane allowing for the local
$\kappa$-symmetry is very complicated. One can hope that our method
being applied to that case, will lead to a more simple formulation.

The work is organized as follows. In Sec. 2 we present and discuss
an action for the auxiliary variable $\Lambda^\mu$, which proves to
be a necessary ingredient of our construction. In Sec. 3 a covariant
action for the eleven dimensional superstring and its local symmetries
are presented. In Sec. 4 within the framework of the Hamiltonian
approach we prove that it has free dynamics. In Sec. 5 the role of the
Wess-Zumino term presented in the action is elucidated. In Sec. 6
off-shell realization of the super Poincar\'e algebra is derived and
discussed. Appendix contains our spinor convention for $D=11$.

\section{Action for auxiliary variables and their dynamics}

As was mentioned in the Introduction, we need to get in our
disposal an auxiliary light-like variable. So, as a preliminary
step of our construction, let us discuss the following $D=11$
Poincar\'e invariant action
\begin{equation}
S=-\int d^2\sigma \left[ \Lambda^\mu\varepsilon^{ab}\partial_a A^\mu_b+
\frac 1\phi \Lambda^\mu\Lambda^\mu\right],
\end{equation}
which turns out to be a building block of the eleven dimensional
superstring action considered below. Here $\Lambda^\mu(\sigma^a)$ is
a $D=11$ vector and a $d2$ scalar, $A^\mu_a(\sigma^b)$ is a $D=11$ and
$d2$ vector, while $\phi(\sigma^a)$ is a scalar field. In Eq. (6) we
have set $\varepsilon^{ab}=-\varepsilon^{ba}$, $\varepsilon^{01}=-1$
and it was also supposed that $\sigma^1\subset [0,\pi]$. From the
equation of motion $\delta S/\delta\phi=0$ it follows that
$\Lambda^\mu$ is a light-like vector.

Local symmetries of the action are $d=2$ reparametrizations\footnote{Note
that the coupling to $d=2$ metric $g^{ab}(\sigma^c)$ is not
necessary due to the presence of the $\varepsilon^{ab}$ symbol and
the supposition that the variable $\phi$ transforms as a density
$\phi'(\sigma')={\rm det}(\partial\sigma'/\partial\sigma)\phi(\sigma)$
under reparametrizations.} and the following transformations with
the parameters $\xi^\mu(\sigma^a)$, $\omega_a(\sigma^b)$:
\begin{eqnarray}
&& \delta_\xi A^\mu_a=\partial_a\xi^\mu;\cr
&& \delta_\omega A^\mu_a=\omega_a\Lambda^\mu, \cr
&& \delta_\omega\phi=\displaystyle\frac 12 \phi^2\varepsilon^{ab}
\partial_a\omega_b.
\end{eqnarray}
These symmetries are reducible because their combination with
the parameters of a special form: $\omega_a=\partial_a\omega$,
$\xi^\mu=-\omega\Lambda^\mu$ is a trivial symmetry: $\delta_\omega
A^\mu_a=-\omega\partial_a\Lambda^\mu$, $\delta_\omega\phi=0$ (note that
$\partial_a\Lambda^\mu=0$ is one of the equations of motion). Thus, Eq.
(7) includes 12 essential parameters which correspond to the primary first
class constraints $p^\mu_0\approx0$, $\pi_\phi\approx0$ in the Hamilton
formalism (see below).

Let us consider the theory in the Hamiltonian framework.
Momenta conjugate to the variables $\Lambda^\mu$, $A^\mu_a$, $\phi$ are
denoted by $p^\mu_\Lambda$, $p^\mu_a$, $\pi_\phi$. All equations for
determining the momenta turn out to be the primary constraints
\begin{eqnarray}
&& \pi_\phi=0, \cr
&& p^\mu_0=0;\\
&& p^\mu_\Lambda=0, \cr
&& p^\mu_1-\Lambda^\mu=0.
\end{eqnarray}
The canonical Hamiltonian is
\begin{equation}
H=\int d\sigma^1\left[\Lambda^\mu\partial_1A^\mu_0+\frac 1\phi \Lambda^2
+\lambda_\phi\pi_\phi+\lambda^\mu_\Lambda p^\mu_\Lambda+\lambda^\mu_0
p^\mu_0+\lambda^\mu_1(p^\mu_1-\Lambda^\mu)\right],
\end{equation}
where $\lambda_*$ are the Lagrange multipliers corresponding to the
constraints. The preservation in time of the primary constraints
implies the secondary ones
\begin{eqnarray}
&& \partial_1\Lambda^\mu=0, \cr
&& \Lambda^2=0,
\end{eqnarray}
and equations for determining some of the Lagrange multipliers
\begin{eqnarray}
&& \lambda^\mu_1=\partial_1 A^\mu_0+\frac 2\phi\Lambda^\mu, \cr
&& \lambda^\mu_\Lambda=0.
\end{eqnarray}
The tertiary constraints are absent.

Constraints (9) form a system of second class and can be omitted
after introducing the corresponding Dirac bracket (the Dirac brackets for
the remaining variables prove to coincide with the Poisson ones).
After imposing the gauge fixing conditions $\phi=2$, $A^\mu_0=0$ for the
first class constraints (8), dynamics of the remaining variables is
governed by the equations
\begin{eqnarray}
&& \dot A^\mu_1=p^\mu_1, \cr
&& \dot p^\mu_1=0;\\
&& (p^\mu_1)^2=0, \cr
&& \partial_1 p^\mu_1=0.
\end{eqnarray}
In order to find a correct gauge for the second constraint in Eq.(14),
let us consider Fourier decomposition of functions periodical in the
interval $\sigma\subset[0,\pi]$
\begin{eqnarray}
&& A^\mu_1(\tau,\sigma)=Y^\mu(\tau)+\sum\limits_{n\ne0} y^\mu_n(\tau)
e^{i2n\sigma},\cr
&& p^\mu_1(\tau,\sigma)=P^\mu_y(\tau)+\sum\limits_{n\ne0}
p^\mu_n(\tau) e^{i2n\sigma}.
\end{eqnarray}
Then the constraint $\partial_1 p^\mu_1=0$ is equivalent to $p^\mu_n=0$,
$n\ne0$, and an appropriate gauge is $y^\mu_n=0$, or, in the equivalent
form $\partial_1 A^\mu_1=0$. Thus, physical degrees of freedom of the
model are the zero modes \footnote{We are grateful to N. Berkovits and
J. Gates for bringing this fact to our attention} of these
variables, and the corresponding dynamics is
\begin{eqnarray}
&& A^\mu_1(\tau,\sigma)=Y^\mu+P^\mu_y\tau,\cr
&& p^\mu_1(\tau,\sigma)=P^\mu_y=\mbox{const}, \cr
&& (P_y)^2=0.
\end{eqnarray}
Since there are no of oscillator variables, the action (6) can be 
considered as describing a point-like object, which propagates freely
according to Eq.(16). The only quantum state is its ground state
$\mid p_{y 0}>$ with the mass $m^2_y=p^2_{y 0}=0$. In the result,
these degrees of freedom do not make contributions into the state
spectrum of the superstring (see Sec.4), and manifest themselves in
additional degeneracy of the continuous part of the energy spectrum
only. The action of such a kind was
successfully used before \cite{26,27} in a different context.

Note that in the previous discussion it was assumed that variables of
the theory are periodical in the interval $\sigma\subset [0,\pi]$.
For an open world sheet, the stationarity condition
$\delta S_\Gamma=0$ for the Hamiltonian action
\[ S_\Gamma=\int d^2\sigma [p_A\dot q^A-H(q,p)] \]
yields
\begin{equation}
\int d\tau \Big(\Lambda^\mu\delta A^\mu_0\big|^{\sigma=\pi}_{\sigma=0}
\Big)=0.
\end{equation}
Since the variations $\delta A^\mu_0|_{\sigma=0,\pi}$ are arbitrary,
this equation requires $\Lambda^\mu|_{\sigma=0,\pi}=0$.
By virtue of Eq. (9) it leads to the trivial solution
$p^\mu_1|_{\sigma=0,\pi}=0$. In contrast, for
the closed world sheet one has $\delta\Gamma^A|_{\sigma=0}=
\delta\Gamma^A|_{\sigma=\pi}$ for any variable $\Gamma^A$ and Eq.
(17) is automatically satisfied. Hence, the model (6) has nontrivial
solution being determined on the closed world sheet only.

\section{Eleven dimensional superstring action and its local
symmetries}

The $D=11$ action functional to be examined is
\begin{eqnarray}
S=\int d^2\sigma\left\{\frac{-g^{ab}}{2\sqrt{-g}}\Pi_a^\mu\Pi^\mu_b-
i\varepsilon^{ab}\partial_a x^\mu(\bar\theta\Gamma^\mu\partial_b
\theta)-\right.\cr
\left.-i\Lambda^\mu\bar\psi\Gamma^\mu\theta-\frac 1\phi \Lambda^\mu
\Lambda^\mu -\Lambda^\mu\varepsilon^{ab}\partial_a A^\mu_b\right\},
\end{eqnarray}
where $\theta$, $\psi$ are 32-component Majorana spinors and
$\Pi^\mu_a\equiv \partial_a x^\mu -i\bar\theta\Gamma^\mu
\partial_a\theta$. Let us mention the origin of the terms presented in
Eq. (18). The first two terms are exactly GS--type superstring action
written in eleven dimensions. The meaning of the last two terms has
been explained in the previous section. The third and the fourth terms
will supply the appearance of the equations $\Lambda_\mu\Gamma_\mu\theta=0$
and $\Lambda^2=0$. Thus, the variables $\bar\psi^\alpha$ and $\phi$
are, in fact, the Lagrange multipliers for these constraints.

Note also that the Wess--Zumino term in the $D=10$ GS action provides
the appearance of the local $\kappa$-symmetry \cite{9}. In our model it
plays a different role, as will be discussed below.

Let us make a comment on the local symmetries structure of the action (18).
Local bosonic symmetries are $d=2$ reparametrizations (with the
standard transformation laws for all variables except for the variable
$\phi$, which transforms as a density: $\phi'(\sigma')={\rm det}(\partial
\sigma'/\partial\sigma)\phi(\sigma)$~), Weyl symmetry, and
the transformations with parameters $\xi^\mu(\sigma^a)$ and
$\omega_a(\sigma^b)$ described in the previous Section.

There is also a fermionic symmetry with parameters
$\chi^\alpha(\sigma^a)$:
\begin{eqnarray}
&& \delta\bar\psi=\bar\chi\Gamma^\mu\Lambda_\mu, \cr
&& \delta\phi=-\phi^2(\bar\chi\theta),
\end{eqnarray}
from which only 16 are essential on-shell since $\Lambda^2=0$. As shown
below, reducibility of this symmetry make no special problem for
covariant quantization.

Let us present arguments that the action constructed describes a free
theory. Equations of motion for the theory (18) are
\begin{mathletters}
\begin{eqnarray}
&& \Pi^\mu_a\Pi^\mu_b-\frac 12 g_{ab}(g^{cd}\Pi^\mu_c\Pi^\mu_d)=0;\\
&& \partial_a\left(\frac{g^{ab}}{\sqrt{-g}}\Pi^\mu_b+i\varepsilon^{ab}
\bar\theta\Gamma^\mu\partial_b\theta\right)=0;\\
&& 4i\Pi^\mu_b(\Gamma^\mu P^{-ba}\partial_a\theta)_\alpha+
\varepsilon^{ab}\theta^\beta\partial_a\theta^\gamma\partial_b
\theta^\delta\Gamma^\mu_{\alpha(\beta}C\Gamma^\mu_{\gamma\delta)}+
i\Lambda^\mu(\Gamma^\mu\psi)_\alpha=0;\\
&& \Lambda^\mu\Gamma^\mu\theta=0, \cr
&& \Lambda^2=0;\\
&& \partial_a\Lambda^\mu=0, \cr
&& \varepsilon^{ab}\partial_a A^\mu_b+
\frac 2\phi\Lambda^\mu+i\bar\psi\Gamma^\mu\theta=0;
\end{eqnarray}
\end{mathletters}
where
\[ P^{-ba}=\frac 12\left(\frac{g^{ba}}{\sqrt{-g}}-\varepsilon^{ba}
\right). \]
Multiplying Eq. (20c) by $\Lambda_\mu\Gamma^\mu$ one gets
\begin{equation}
(\Lambda^\mu\Pi^\mu_b)P^{-ba}\partial_a\theta=0.
\end{equation}
In the coordinate system where $\Lambda^-=1$, supplemented by the
conformal gauge, it can be rewritten as
\begin{equation}
(\partial_0+\partial_1)\theta=0,
\end{equation}
from which it follows that any solution $\theta(\sigma)$ of the system
(20) obeys this free equation.

Thus, Eqs. (20a--c) for the $g^{ab}$, $x^\mu$, $\theta^\alpha$ variables
in fact coincide with those of the GS string and are accompanied by
$\Lambda_\mu\Gamma^\mu\theta=0$. The latter reduces to $\Gamma^+\theta=0$
in the coordinate system chosen. In the result, one expects free
dynamics in this sector provided that the conformal gauge has been
assumed.  In the next section we will rigorously prove this fact by
direct calculations in the Hamiltonian framework.

\section{Analysis of dynamics}

From the explicit form of the action functional (18) it follows that the
variable $\Lambda^\mu$ can be excluded by making use of its equation of
motion. The Hamiltonian analog of the situation is a pair of second
class constraints ${p_\Lambda}^\mu=0$, ${p_1}^\mu-\Lambda^\mu=0$,
which can be omitted after introducing the associated Dirac bracket
(see Sec. 2). The Dirac brackets for the remaining variables prove to
coincide with the Poisson ones and the Hamiltonian looks like
\begin{eqnarray}
& H=\displaystyle\int d\sigma^1\left\{-\frac N2(\hat p^2+\Pi_{1\mu}
\Pi_1^\mu)-N_1\hat p_\mu \Pi_1^\mu +p_{1\mu}(\partial_1 A_0^\mu+
i\bar\psi\Gamma^\mu\theta)+\right. \cr
& \left.+\displaystyle\frac 1\phi (p_1^\mu)^2+ \lambda_\phi\pi_\phi +
\lambda_{0\mu}p_0^\mu +\lambda^{ab}(\pi_g)_{ab}+{\lambda_\psi}^\alpha
p_{\psi\alpha} +L_\alpha{\lambda_\theta}^\alpha\right\},\label{ham}
\end{eqnarray}
where $p^\mu$, $p_0^\mu$, $p_1^\mu$, $p_{\psi\alpha}$, $(\pi_g)_{ab}$
are momenta conjugate to the variables $x^\mu$, $A_0^\mu$, $A_1^\mu$,
$\psi_\alpha$, $g_{ab}$, respectively; $\lambda_*$ are Lagrange
multipliers corresponding to the primary constraints. In Eq. (23) we
also denoted
\begin{eqnarray}
&& N = \frac{\sqrt{-g}}{g^{00}}, \cr
&& N_1=\frac{g^{01}}{g^{00}}, \cr
&& \hat p^\mu=p^\mu-i\bar\theta\Gamma^\mu \partial_1\theta, \cr
&& L_\alpha\equiv p_{\theta\alpha}-i(p^\mu+\Pi_1^\mu)(\bar\theta
\Gamma^\mu)_\alpha=0.
\end{eqnarray}

It is interesting to note that the fermionic constraints $L_\alpha=0$
obey the algebra (2) and, being considered on their own (without taking
into account the constraints $\bar\theta \Gamma^\mu p_{1\mu}=0$
which will arise below), form a system which has no definite class
(this corresponds to the lack of $\kappa$-symmetry in the GS action
written in eleven dimensions).

The conservation in time of the primary constraints implies the
secondary ones
\begin{eqnarray}
&& \partial_1 p^\mu_1=0, \cr
&& (p^\mu_1)^2=0, \cr
&& (\bar\theta\Gamma^\mu)_\alpha p^\mu_1=0,\cr
&& (\hat p^\mu\pm\Pi^\mu_1)^2=0;
\end{eqnarray}
\begin{eqnarray}
&& (\bar\lambda_\theta\Gamma^\mu)_\alpha(\hat p^\mu+\Pi^\mu_1)
+i\bar\theta^\gamma\partial_1\theta^\delta\lambda^\beta_\theta
\Gamma^\mu_{\gamma(\delta}C\Gamma^\mu_{\beta\alpha)}+
\frac 12(\bar\psi\Gamma^\mu)_\alpha\Lambda^\mu-\cr
&& -(\partial_1\bar\theta\Gamma^\mu)_\alpha(N+N_1)(\hat p^\mu+\Pi^\mu_1)
-\frac 12(\bar\theta\Gamma^\mu)_\alpha\partial_1
(N\hat p^\mu+N_1\Pi^\mu_1)=0.
\end{eqnarray}
At the next step, there arises only one nontrivial equation. From
the condition $\{\bar\theta\Gamma^\mu p^\mu_1,H\}=0$ one gets
\begin{equation}
(\bar\lambda_\theta\Gamma^\mu)_\alpha p^\mu_1=0.
\end{equation}
Equations (25), (26) are equivalent to
\begin{eqnarray}
&& \bar\lambda_\theta=(N+N_1)\partial_1\bar\theta+
\displaystyle\frac{\tilde\xi}2\bar\theta,\\
&& \tilde S_\alpha\equiv (\bar\psi\Gamma^\mu)_\alpha p^\mu_1+
(\bar\theta\Gamma^\mu)_\alpha\tilde D^\mu=0,
\end{eqnarray}
where we denoted
\begin{eqnarray*}
&& \tilde D^\mu=\tilde\xi(\hat p^\mu+\Pi^\mu_1)-\partial_1(N\hat p^\mu+
N_1\Pi^\mu_1), \\
&& \tilde\xi=\frac{\partial_1(N\hat p^\mu+N_1\Pi^\mu_1)
p^\mu_1}{(\hat p^\mu+\Pi^\mu_1)p^\mu_1}.
\end{eqnarray*}
Thus, we have Eq. (27) for determining the Lagrange multiplier
$\lambda_\theta$ and the tertiary constraint $\tilde S_\alpha=0$. One can
check that there are no more constraints in the problem.

Hamiltonian equations of motion for the variables $(g^{ab},(\pi_g)_{ab})$,
$(\phi,\pi_\phi)$, $(A^\mu_0,p^\mu_0)$, $(\psi^\alpha,p^\alpha_\psi)$
look as follows: $\partial_0q=\lambda_q$, $\partial_0p_q=0$,
while for other variables one has
\begin{mathletters}
\begin{eqnarray}
&& \partial_0A^\mu_1=\partial_1A^\mu_0+\displaystyle\frac 2\phi p^\mu_1+
i\bar\psi\Gamma^\mu\theta, \cr
&& \partial_0 p^\mu_1=0,\\
&& \partial_0 x^\mu=-N\hat p^\mu-N_1\Pi^\mu_1-i\bar\theta\Gamma^\mu
\lambda_\theta, \cr
&& \partial_0p^\mu=-\partial_1(N\Pi^\mu_1+
N_1 \hat p^\mu)+i\bar\theta\Gamma^\mu\lambda_\theta,\\
&& \partial_0\theta^\alpha=-\lambda^\alpha_\theta.
\end{eqnarray}
\end{mathletters}
Note that equations $\partial_0p_{\theta\alpha}=\dots$ have been omitted
since they follow from the constraints $L_\alpha=0$ and other
equations.

To go further, note that the constraints $(\pi_g)_{ab}=0$ form a
nonvanishing Poisson bracket with the $\tilde S_\alpha$ from Eq. (29). A
modification which splits them out of other constraints is
\[ (\tilde\pi_g)_{ab}\equiv (\pi_g)_{ab}+\frac 1{2(\hat p+\Pi_1)p_1}
(p_\psi\Gamma^\mu \Gamma^\nu\theta)(\hat p^\mu+\Pi_1^\mu){T^\nu}_{ab}, \]
with $T^\nu_{ab}$ being defined by the equality
$\{(\pi_g)_{ab},\tilde S_\alpha\}={T^\mu}_{ab}(\bar\theta\Gamma^\mu)_\alpha$.
Hence, the constraints $(\tilde \pi_g)_{ab}=0$ are first class and one can
adopt the gauge choice $g^{ab}=\eta^{ab}$. The full set of constraints
can now be rewritten in a more simple form
\begin{mathletters}
\begin{eqnarray}
&& \pi_\phi=0, \cr
&& p_0^\mu=0;\\
&& (p_1^\mu)^2=0, \cr
&& \partial_1p_1^\mu=0, \cr
&& (\hat p^\mu \pm \Pi_1^\mu)^2=0,\cr
&& L_\alpha=0, \cr
&& \bar\theta\Gamma^\mu p_{1\mu}=0,\cr
&& p_{\psi\alpha}=0, \cr
&& S_\alpha\equiv \bar\psi \Gamma^\mu p_{1\mu}
+ (\bar\theta \Gamma^\mu)_\alpha D_\mu=0.
\end{eqnarray}
\end{mathletters}
where
\begin{eqnarray}
&& D^\mu\equiv \xi(\hat p^\mu+\Pi_1^\mu)-\partial_1 p^\mu, \cr
&& \xi\equiv \frac{\partial_1 \hat p^\mu p_{1\mu}}{(\hat p^\nu
+\Pi_1^\nu) p_{1\nu}}.
\end{eqnarray}

Now, let us impose gauge fixing conditions to the first class
constraints (30.a). The choice consistent with the equations of motion
is
\begin{eqnarray*}
&& \phi=2, \\
&& A_0^\mu=-i\int_0^\sigma d\sigma' \bar\psi \Gamma^\mu\theta.
\end{eqnarray*}
After that, dynamics for the remaining variables looks like
\begin{mathletters}
\begin{eqnarray}
&& \partial_0\psi^\alpha={\lambda_\psi}^\alpha,\cr
&& \partial_0 p_{\psi\alpha}=0,\cr
&& p_{\psi\alpha}=0, \cr
&& S_\alpha=0;\\
&& \partial_0 A_1^\mu=p_1^\mu,\cr
&& \partial_0 p_1^\mu=0,\cr
&& (p_1^\mu)^2=0, \cr
&& \partial_1 p_1^\mu=0;\\
&& \partial_0 x^\mu=-p^\mu,\cr
&& \partial_0 p^\mu=-\partial_1\partial_1 x^\mu,\cr
&& (\hat p^\mu\pm\Pi_1^\mu)^2=0;\\
&& \partial_0\theta=-\partial_1\theta-\displaystyle\frac \xi 2\theta,\cr
&& L_\alpha=0, \cr
&& (\bar\theta \Gamma^\mu)_\alpha p_{1\mu}=0.
\end{eqnarray}
\end{mathletters}

The sector (33.a) includes $ 32+16$ independent constraints from which
the first class ones can be picked out as follows:
\begin{equation}
(p_\psi \Gamma^\mu)_\alpha p_{1\mu}=0.
\end{equation}
As was mentioned above, reducibility of the constraints does not spoil
the covariant quantization program. Actually, let us impose the following
covariant (and redundant) gauge fixing conditions for the constraints
(34):
\begin{equation} {S^1}_\alpha \equiv \frac 1{(\hat p+\Pi_1)p_1}
\bar\psi \Gamma^\mu (\hat p_\mu+\Pi_{1\mu})=0.
\end{equation}
Then the set of equations $S_\alpha=0$, ${S^1}_\alpha=0$ is equivalent to
\begin{equation}
S' \equiv \bar\psi -\frac 1{2(\hat p+\Pi_1)p_1} \bar\theta
\Gamma^\mu D_\mu \Gamma^\nu (\hat p_\nu +\Pi_{1\nu}),
\end{equation}
the latter forms a nondegenerate Poisson bracket together with the
constraints $p_{\psi\alpha}=0$
\begin{equation}
\{ p_{\psi\alpha}, S'_\beta\}=-C_{\alpha\beta}.
\end{equation}
After passing to the Dirac bracket associated with the second class
functions $p_{\psi\alpha}$, $S'_\alpha$, the variables $\psi$,
$p_\psi$ can be dropped.

To proceed further, we impose the gauge $\partial_1 A_1^\mu=0$ for
the constraints in Eq. (33b), and pass to an appropriately chosen
coordinate system. By making use of the Lorentz transformation one can
consider a coordinate system where $P^\mu_y=(1,0,\dots,0,1)$ (note, that
it is admissible procedure within the canonical quantization approach
since the Lorentz transformation is a particular example a canonical
one). To get dynamics in the final form, we pass to the light-cone
coordinates $x^\mu \to (x^+,x^-,x^i)$, $i=1,2,\dots,8,10$,
$\theta^\alpha \to (\theta_a, \bar\theta'_{\dot a}, \theta'_a,
\bar\theta_{\dot a})$, $a,\dot a=1,\dots,8$ and impose the gauge fixing
conditions
\begin{eqnarray}
&& x^+=P^+\tau, \cr
&& p^+=-P^+={\rm const},
\end{eqnarray}
to the Virasoro first class constraints remaining in Eq.(33c).
The equation $\bar\theta\Gamma^\mu p_{1\mu}=0$ acquires now the form
$\Gamma^+\theta=0$ and it is easy to show that $32+16$ constraints
$L_\alpha=0$, $\Gamma^+\theta=0$ are second class.
A solution is $\theta^\alpha=(\theta_a,0,0,\bar\theta_{\dot a})$
with $\theta_a$ and $\bar\theta_{\dot a}$ being $SO(8)$ spinors of
opposite chirality. In the gauge chosen, the
relation $(\hat p^\mu+\Pi_1^\mu)p_{1\mu}\ne0$ holds which correlates
with the assumption made above in Eqs. (32), (35). For the remaining
variables one gets free field equations
\begin{eqnarray}
&& \partial_0 x^i=-p^i, \cr
&& \partial_0 p^i=-\partial_1 \partial_1 x^i;\cr
&& (\partial_0+\partial_1)\theta_a=0, \cr
&& (\partial_0+\partial_1)\bar\theta_{\dot a}=0.
\end{eqnarray}
Moreover, $\theta_a$ and $\bar\theta_{\dot a}$ form two pairs of
selfconjugate variables under the Dirac bracket associated with the
constraints from Eq. (33d)
\begin{eqnarray}
&& \{\theta_a,\theta_b\}=\frac i{\sqrt 8 P^+}\delta_{ab}, \cr
&& \{\bar\theta_{\dot a},\bar\theta_{\dot b}\}=\frac i{\sqrt 8 P^+}
\delta_{\dot a\dot b}.
\end{eqnarray}
Let us look shortly at the spectrum of the theory. The ground state of
the full theory 
$\mid p_{y 0}, p_0, 0 >= \mid p_{y 0} > \mid p_0 > \mid 0 >$ is a 
direct product of vacua, where $P^2_y\mid p_{y 0} >=0$,
$\mid p_0 >$ is a vacuum for zero modes of the variables 
$x^\mu, p^\mu$, while through $\mid 0 >$ are denoted vacua 
for bosonic and
fermionic oscillator modes. From Eq.(40) it follows that zero modes of
the $\theta_a, \bar\theta_{\dot a}$ variables form the Clifford
algebra which is also symmetry algebra of a ground state. A
representation space is 256-dimensional which corresponds to
the spectrum of the $D=11$ supergravity \cite{29}. The excitation 
levels are then obtained by acting with oscillators on the ground 
state. One notes that zero modes $Y^\mu, P^\mu_y$ manifest 
themselves in additional degeneracy of the continuous energy spectrum
only.

\section{A comment on the Wess--Zumino term in the $D=11$
superstring action}

For the $D=10$ GS superstring the Wess-Zumino term provides the local
$\kappa$-symmetry \cite{11,12}, which leads to free dynamics for
physical variables. Since there is no $\kappa$-symmetry in our
construction, it is interesting to elucidate the meaning of this term
in the $D=11$ action suggested. Let us consider the action (18) with
the second term omitted. Canonical analysis for this model turns out to
be very similar to that made above and we present results only.

Instead of Eqs. (24), (28), (29) one finds
\begin{eqnarray}
&& L_\alpha\equiv p_{\theta\alpha}-i(\bar\theta\Gamma^\mu)_\alpha
p^\mu=0,\cr
&& \bar\lambda_\theta=\displaystyle\frac{N(\Pi_1p_1)+N_1(pp_1)}
{(pp_1)}\partial_1\bar\theta,\\
&& \tilde S\equiv [(pp_1)\bar\psi-\partial_1\bar\theta\Gamma^\rho
(N\Pi^\rho_1+N_1p^\rho)\Gamma^\nu p^\nu]\Gamma^\mu p^\mu_1=0.\nonumber
\end{eqnarray}
In the coordinate system where $P^\mu_y=(1,0,\dots,0,1)$
the analog of the equations (33c), (33d) reads
\begin{eqnarray}
&& \partial_0x^\mu=-p^\mu-i\displaystyle\frac{\partial_1x^+}{p^+}
(\bar\theta\Gamma^\mu\partial_1\theta), \cr
&& \partial_0p^\mu=-\partial_1\Pi^\mu_1,\cr
&& (p^\mu\pm\Pi^\mu_1)^2=0;\cr
&& \partial_0\theta=-\frac{\partial_1x^+}{p^+}\partial_1\theta,\cr
&& L_\alpha=0, \cr
&& \Gamma^+\theta=0,
\end{eqnarray}
provided that the conformal gauge has been chosen.

To impose a gauge for the first class constraints $(p^\mu\pm\Pi^\mu_1)^2=0$,
consider one-parameter set of equations\footnote{The value $c=\pm1$ is
not admissible since in that case the Poisson bracket of the
constraints $(p^\mu\pm\Pi^\mu_1)^2=0$ and the gauges (43) vanishes.}
\begin{eqnarray}
&& x^+=P^+(\tau+c\sigma), \cr
&& p^+=-P^+={\rm const}, \cr
&& c={\rm const}\ne \pm1,
\end{eqnarray}
which leads to the following dynamics for variables of the physical sector:
\begin{eqnarray}
&& \partial_0x^i=-p^i, \cr
&& \partial_0p^i=-\partial_1\partial_1x^i,\cr
&& (\partial_0-c\partial_1)\theta=0.
\end{eqnarray}
One can check that it is impossible to get rid of the number $c$ by
making use of some other gauge choice for $g^{ab}$ and $A^\mu_1$ variables.

Thus, omitting the Wess--Zumino term in Eq. (18) one arrives at the
theory which possesses all the properties of the model (18) with the
only modification in the last of Eqs. (39): $(\partial_0-c\partial_1)
\theta=0$ with $c$ a constant. Depending on the gauge chosen it can take any
value except $c=\pm1$. Hence, the dynamics is not manifestly $d=2$ Poincar\'e
covariant, provided that $\theta$ be a $d=2$ scalar. It is the
Wess--Zumino term which corrects this inconsistency.

\section{Off-shell realization of the $D=11$ super-Poincar\'e
algebra}

It is convenient first to recall some facts relating to the $D=10$ GS
superstring. Off-shell realization of the super
Poincar\'e algebra for that case includes the Poincar\'e transformations
accompanied by the supersymmetries
\begin{eqnarray}
&& \delta\theta^\alpha=\epsilon^\alpha, \cr
&& \delta x^\mu=-i\bar\theta\Gamma^\mu\epsilon.
\end{eqnarray}
Being considered on their own, in the gauge $\Gamma^+\theta=0$ these
transformations are reduced to trivial shifts for variables of the
physical sector
\begin{eqnarray}
&& \delta\bar\theta_{\dot a}=\bar\epsilon_{\dot a}, \cr
&& \delta x^i=0.
\end{eqnarray}
To get on-shell realization of the supersymmetry algebra, one needs to
consider a combination of the $\epsilon$- and $\kappa$-transformations
$\delta_\epsilon+\delta_{\kappa(\epsilon)}$, which does not violate the
gauge $\Gamma^+\theta=0$. These transformations are (see, for example,
Ref. 28)
\begin{eqnarray}
&& \delta\bar\theta_{\dot a}=\bar\epsilon_{\dot a}+\frac 1{P^+}
\partial_- x^i \bar\gamma^i{}_{\dot aa}\epsilon_a, \cr
&& \delta x^i=-i\sqrt{2}(\bar\theta\bar\gamma^i\epsilon).
\end{eqnarray}

We turn now to the $D=11$ case. Off-shell realization of the super
Poincar\'e algebra for the action (18) includes the Poincar\'e
transformations in the standard realization and the following
supersymmetries with 32-component spinor parameter $\epsilon^\alpha$:
\begin{eqnarray}
&& \delta\theta=\tilde\Lambda\epsilon, \cr
&& \delta x^\mu=-i\bar\theta\Gamma^\mu\tilde\Lambda\epsilon,\cr
&& \delta {A^\mu}_a=-2i\epsilon_{ab}\displaystyle\frac{g^{bc}}{\sqrt{-g}}
(\bar\theta\tilde\Pi_c{\Gamma^\mu}\epsilon)-
2i\partial_ax^\nu(\bar\theta\Gamma^\nu\Gamma^\mu\epsilon)-
2(\bar\theta\epsilon)(\bar\theta\Gamma^\mu\partial_a\theta),\\
&& \delta\bar\psi=i\epsilon^{ab}[\bar\epsilon\Gamma^\mu(\partial_a
\bar\theta\Gamma^\mu\partial_b\theta)-2\partial_a\bar\theta
(\partial_b\bar\theta\epsilon)],\cr
&& \delta\phi=-i\phi^2(\bar\psi\epsilon),\nonumber
\end{eqnarray}
where $\tilde\Lambda\equiv\Lambda_\mu\Gamma^\mu$, $\tilde\Pi_c\equiv
{\Pi_c}^\mu\Gamma^\mu$. The action is invariant up to total derivative
terms. These transformations are the analog of Eq. (45) since in the
physical sector they are reduced to $\delta\theta_a=\sqrt2
\epsilon'_a$, $\delta\bar\theta_{\dot a}=-\sqrt2 \bar\epsilon'_{\dot
a}$, $\delta x^i=0$.

To find a global supersymmetry of the action (18) corresponding
to Eq. (47) let us consider the following ansatz:
\begin{eqnarray}
&& \delta\theta=\tilde\Lambda\tilde\Pi_c\epsilon^c, \cr
&& \delta\phi=-i\phi^2(\bar\psi\tilde\Pi_c\epsilon^c),\\
&& \delta x^\mu=4i(\Lambda\Pi_c)(\bar\theta\Gamma^\mu\epsilon^c)+
2i(\bar\theta\tilde\Pi_c\epsilon^c)\Lambda^\mu,\nonumber
\end{eqnarray}
where we denoted
\begin{eqnarray}
&& \epsilon^a_\alpha\equiv P^{-ab}\epsilon_{\alpha\,b}, \cr
&& P^{-ab}=\frac 12\left(\frac{g^{ab}}{\sqrt{-g}}-\varepsilon^{ab}\right), \\
&& (\Lambda\Pi_c)\equiv\Lambda^\mu{\Pi_c}^\mu.\nonumber
\end{eqnarray}
Variation of the GS part of the action (18) under these transformations
looks like
\begin{eqnarray}
\lefteqn{\delta S_{GS}=\varepsilon^{ab}[-8(\bar\theta\Gamma^\mu\epsilon^c)
(\partial_a\bar\theta\Gamma^\mu\partial_b\theta)(\Lambda\Pi_c)-
4(\bar\theta\tilde\Pi_c\epsilon^c)(\partial_a\bar\theta
\tilde\Lambda\partial_b\theta)+}\cr
&& +2(\partial_a\bar\theta\Gamma^\mu \tilde\Lambda\tilde\Pi_c\epsilon^c)
(\bar\theta\Gamma^\mu\partial_b \theta)+(\bar\theta\Gamma^\mu
\tilde\Lambda\tilde\Pi_c\epsilon^c)(\partial_a\bar\theta\Gamma^\mu
\partial_b\theta)]-\cr
&& -2iP^{-ba}[4(\bar\theta\tilde\Pi_c\epsilon^c)(\partial_a\Lambda
\Pi_b)+2(\partial_a\bar\theta\tilde\Lambda\epsilon^c)(\Pi_b\Pi_c)-
(\bar\theta\tilde\Lambda\partial_a\tilde\Pi_b\tilde\Pi_c
\epsilon^c)].
\end{eqnarray}
After integrating by parts, reordering the $\tilde\Lambda$
and $\tilde\Pi$ terms and making use of the identities
\begin{eqnarray}
&& P^{-ab}P^{-cd}=P^{-cb}P^{-ad},\cr
&& (\partial_a\bar\theta\Gamma^\mu\partial_b\theta)(\Lambda\Pi_c)=
-\frac 12 \partial_a\bar\theta\Gamma^\mu\{\tilde\Lambda,
\tilde\Pi_c\}\partial_b\theta,
\end{eqnarray}
it proves to be possible to represent all the terms in Eq. (51) either
as $K\tilde\Lambda\theta$ or $\partial_a\Lambda^\mu T^{\mu a}$ with $K$
and $T$ being certain coefficient. These terms can evidently be
canceled by appropriate variations of the $\bar\psi$ and $A_\mu^a$
variables.  The final form for these variations is \begin{eqnarray}
\lefteqn{\delta A^\mu_a=8(\bar\theta\Gamma^\rho\epsilon^c)
(\bar\theta\Gamma^\mu\Pi^\nu_c\Gamma^{\nu\rho}\partial_a\theta)-
5(\bar\theta\tilde\Pi_c\epsilon^c)(\bar\theta\Gamma^\mu
\partial_a\theta)-}\cr
&& -3\bar\theta\Gamma^\mu\Gamma^\nu\tilde\Pi_c\epsilon^c)
(\bar\theta\Gamma^\nu\partial_a\theta)-
4i\varepsilon_{ad}P^{-bd}[(\bar\theta\Gamma^\mu\epsilon^c)(\Pi_b\Pi_c)-\cr
&& -2(\bar\theta\tilde\Pi_c\epsilon^c)\Pi^\mu_b],\\[1ex]
\lefteqn{\delta\bar\psi=i\varepsilon^{ab}\{2(\partial_a\bar\theta
\tilde\Pi_c\epsilon^c)\partial_b\bar\theta-8(\partial_a\bar\theta
\tilde\Pi_c\partial_b\theta)\bar\epsilon^c-
8\partial_a[(\bar\theta\Gamma^\mu\epsilon^c)\partial_b\bar\theta
\Gamma^{\mu\nu}\Pi_\nu^c]+}\cr
&& +5(\bar\theta\partial_a\tilde\Pi_c\epsilon^c)\partial_b\theta
+3(\bar\theta\Gamma^\mu\partial_b\theta)\bar\epsilon^c\partial_a
\tilde\Pi_c\Gamma^\mu+(\partial_a\bar\theta\Gamma^\mu\partial_b\theta)
\bar\epsilon^c\tilde\Pi_c\Gamma^\mu\}-\cr
&& -2iP^{-ba}[\bar\epsilon^c\partial_a\tilde\Pi_c\tilde\Pi_b-
2\bar\epsilon^c\Pi_b\partial_a\Pi_c].\nonumber
\end{eqnarray}
Note that the complicated transformation law for the $\psi$-variable
might be predicted, since one of the Lagrangian equations of motion is
\begin{equation}
(\tilde\Lambda\psi)_\alpha=-4\tilde\Pi_b P^{-ba}\partial_a\theta_\alpha
+i\varepsilon^{ab}\theta^\beta\partial_a\theta^\gamma\partial_b
\theta^\delta\Gamma^\mu_{\alpha(\beta}(C\Gamma^\mu)_{\gamma\delta)}.
\end{equation}
Thus, transformation of the $\tilde\Lambda\psi$ part of the
$\psi$-variable is dictated by this equation and the transformation laws
for the $x$ and $\theta$ variables.

Being reduced to the physical sector, Eq. (49) looks as follows:
\begin{eqnarray}
&& \delta\theta_a=-\sqrt2 (P^+\epsilon_a-\partial_-x^i{\gamma^i}_{a\dot
a}\bar\epsilon'_{\dot a}+\partial_-x^{10}\epsilon'_a),\cr
&& \delta\bar\theta_{\dot a}=-\sqrt2 (P^+\bar\epsilon_{\dot a}
+\partial_-x^i\bar\gamma^i_{\dot aa}\epsilon'_a-
\partial_-x^{10}\bar\epsilon'_{\dot a}),\\
&& \delta x^i=2\sqrt 2 iP^+(\theta\gamma^i\bar\epsilon'-
\bar\theta\bar\gamma^i\epsilon')\nonumber
\end{eqnarray}
and seems to be the analog of Eq. (47).

To summarize, in this paper we have suggested a super Poincar\'e
invariant action for the superstring which classically exists in any
spacetime dimension. As compare with GS formulation for $N=1, D=10$
superstring action, the only difference is an additional infinite
degeneracy in the continuous part of the energy spectrum, related
with the zero modes $Y^\mu, P^\mu_y$. Since supersymmetry is realised
in the physical subspace (55), one also gets the corresponding
representation in the space of functions on that subspace. This allows
one to expect a supersymmetric spectrum of quantum states.
Analysis of this situation in terms of oscillator variables
as well as the critical dimension will be presented in a separate
publication.

The authors are grateful to N. Berkovits, I.L. Buchbinder, J. Gates
and D.M. Gitman for useful discussions. This work was supported 
by Joint DFG-RFBR project No 96-02-00180G, INTAS Grant-96-0308
(A.D.), INTAS-RFBR Grant No 95-829 (A.G.) and by FAPESP (A.D. and
A.G.).

{\bf Note added:} After this work has been completed,
there appeared a paper by Bars and Deliduman \cite{30} where  a
covariant action for superstring in a space with the nonstandard
signature $(D-2,2)$ was suggested.

\appendix
\section{}
In this Appendix we describe the minimal spinor representation of the
Lorentz group $SO(1,10)$ which is know to have dimension $2^{[D/2]}$.
For this aim, it suffices to find eleven $32\times32$ $\Gamma^\mu$-matrices
satisfying the equation $\Gamma^\mu\Gamma^\nu+\Gamma^\nu\Gamma^\mu=
-2\eta^{\mu\nu}$, $\mu,\nu=0,1,\dots,10$, $\eta^{\mu\nu}=(+,-,\dots,-)$.
A convenient way is to use the well known $16\times 16$
$\Gamma$-matrices of $SO(1,9)$ group which we denote as
$\Gamma^m_{\alpha\beta}$, $\tilde\Gamma^{m\alpha\beta}$, $m=0,1,\dots,9$.
Their explicit form is:
\begin{eqnarray}
&& \Gamma^0=\left(\begin{array}{cc} {\bf 1}_8 & 0\\
0 & {\bf 1}_8\end{array}\right), \cr
&& \tilde\Gamma^0=\left(\begin{array}{cc} -{\bf 1}_8 & 0\\
0 & -{\bf 1}_8\end{array}\right),\cr
&& \Gamma^i=\left(\begin{array}{cc} 0 & {\gamma^i}_{a\dot a}\\
\bar\gamma^i{}_{\dot aa} & 0 \end{array}\right), \cr
&& \tilde\Gamma^i=\left(\begin{array}{cc} 0 & {\gamma^i}_{a\dot a}\\
\tilde\gamma^i{}_{\dot aa} & 0\end{array}\right),\cr
&& \Gamma^9=\left(\begin{array}{cc} {\bf 1}_8 & 0\\
0 & -{\bf 1}_8\end{array}\right), \cr
&& \tilde\Gamma^9=\left(\begin{array}{cc} {\bf 1}_8 & 0\\
0 & -{\bf 1}_8\end{array}\right),
\end{eqnarray}
where ${\gamma^i}_{a\dot a}$, $\bar\gamma^i{}_{\dot aa}\equiv
({\gamma^i}_{a\dot a})^{\rm T}$ are real $SO(8)$ $\gamma$-matrices
\cite{29}
\begin{equation}
\gamma^i\bar\gamma^j+\gamma^j\bar\gamma^i=2\delta^{ij}{\bf 1}_8,
\end{equation}
where $i,a,\dot a=1,\dots,8$. As a consequence, the matrices $\Gamma^m$,
$\tilde\Gamma^m$ are real, symmetric and obey the algebra
\begin{equation}
\{\Gamma^m, \tilde\Gamma^n\}=-2\eta^{mn}{\bf 1},
\end{equation}
where $\eta^{mn}=(+,-,\dots,-)$. Then a possible realization for the $D=11$
$\Gamma$-matrices is
\begin{equation}
\Gamma^\mu=\left\{\left(\begin{array}{cc} 0 & \Gamma^m\\
\tilde\Gamma^m & 0\end{array}\right), \left(\begin{array}{cc}
{\bf 1}_{16} & 0\\ 0 & -{\bf 1}_{16}\end{array}\right)\right\},
\end{equation}
where $\mu=0,1,\dots,10$. The properties of $\Gamma^m$,
$\tilde\Gamma^m$ induce the following relations for $\Gamma^\mu$:
\begin{eqnarray}
&& (\Gamma^0)^{\rm T}=-\Gamma^0, \cr
&& (\Gamma^i)^{\rm T}=-\Gamma^i, \cr
&& (\Gamma^\mu)^*=\Gamma^\mu,\cr
&& \{\Gamma^\mu,\Gamma^\nu\}=-2\eta^{\mu\nu}{\bf 1}_{32},
\end{eqnarray}
where $\eta^{\mu\nu}=(+,-,\dots,-)$. The charge conjugation matrix
\begin{eqnarray}
&& C\equiv\Gamma^0, \cr
&& C^{-1}=-C, \\
&& C^2=-{\bf 1}\nonumber
\end{eqnarray}
can be used to construct the symmetric matrices: $(C\Gamma^\mu)^{\rm
T}=C\Gamma^\mu$.

The next step is to introduce the antisymmetrized products
\begin{equation}
\Gamma^{\mu\nu}=\frac 12(\Gamma^\mu\Gamma^\nu-\Gamma^\nu\Gamma^\mu),
\end{equation}
which have the following explicit form in terms of the corresponding
$SO(1,9)$ and $SO(8)$ matrices:
\begin{eqnarray}
&& \Gamma^{0i}=\left(\begin{array}{cc}
\Gamma^{0i} & 0\\ 0 & \tilde\Gamma^{0i}\end{array}\right)=\left(
\mbox{\begin{tabular}{c|c}
$\begin{array}{cc} 0 & \gamma^i\\ \bar\gamma^i & 0\end{array}$ & 0\\
\hline
0 & $\begin{array}{cc} 0 & -\gamma^i\\ -\bar\gamma^i & 0\end{array}$
\end{tabular}}\right),\cr
&& \Gamma^{09}=\left(\begin{array}{cc}
\Gamma^{09} & 0\\ 0 & \tilde\Gamma^{09}\end{array}\right)=\left(
\mbox{\begin{tabular}{c|c}
$\begin{array}{cc} 1 & 0\\ 0 & -1\end{array}$ & 0\\
\hline
0 & $\begin{array}{cc} -1 & 0\\ 0 & 1\end{array}$
\end{tabular}}\right),\cr
&& \Gamma^{ij}=\left(\begin{array}{cc}
\Gamma^{ij} & 0\\ 0 & \tilde\Gamma^{ij}\end{array}\right)=\left(
\mbox{\begin{tabular}{c|c}
$\begin{array}{cc} \gamma^{ij} & 0\\ 0 & \bar\gamma^{ij}\end{array}$ & 0\\
\hline
0 & $\begin{array}{cc} \gamma^{ij} & 0\\ 0 & \bar\gamma^{ij}\end{array}$
\end{tabular}}\right),\cr
&& \Gamma^{i9}=\left(\begin{array}{cc}
\Gamma^{i9} & 0\\ 0 & \tilde\Gamma^{i9}\end{array}\right)=\left(
\mbox{\begin{tabular}{c|c}
$\begin{array}{cc} 0 & -\gamma^i\\ \bar\gamma^i & 0\end{array}$ & 0\\
\hline
0 & $\begin{array}{cc} 0 & -\gamma^i\\ \bar\gamma^i & 0\end{array}$
\end{tabular}}\right),\\
&& \Gamma^{0,10}=\left(\begin{array}{cc}
0 & -\Gamma^0\\ \tilde\Gamma^0 & 0\end{array}\right)=\left(
\mbox{\begin{tabular}{c|c}
0 & $\begin{array}{cc} {\bf 1} & 0\\ 0 & {\bf 1}\end{array}$\\
\hline
$\begin{array}{cc} {\bf 1} & 0\\ 0 & {\bf 1}\end{array}$ & 0
\end{tabular}}\right),\cr
&& \Gamma^{i,10}=\left(\begin{array}{cc}
0 & -\Gamma^i\\ \tilde\Gamma^i & 0\end{array}\right)=\left(
\mbox{\begin{tabular}{c|c}
0 & $\begin{array}{cc} 0 & -\gamma^i\\ -\bar\gamma^i & 0\end{array}$\\
\hline
$\begin{array}{cc} 0 & \gamma^i\\ \bar\gamma^i & 0\end{array}$ & 0
\end{tabular}}\right),\cr
&& \Gamma^{9,10}=\left(\begin{array}{cc}
0 & -\Gamma^9\\ \tilde\Gamma^9 & 0\end{array}\right)=\left(
\mbox{\begin{tabular}{c|c}
0 & $\begin{array}{cc} -{\bf 1} & 0\\ 0 & {\bf 1}\end{array}$\\
\hline
$\begin{array}{cc} {\bf 1} & 0\\ 0 & -{\bf 1}\end{array}$ & 0
\end{tabular}}\right),
\end{eqnarray}
where $i=1,2,\dots,8$ and $\Gamma^{0i}$, $\Gamma^{09}$, $\Gamma^{0,10}$
are symmetric, whereas $\Gamma^{ij}$, $\Gamma^{i9}$, $\Gamma^{i,10}$,
$\Gamma^{9,10}$ are antisymmetric. Besides, these matrices are real and,
as a consequence of Eq. (A5), obey the commutation relations of the
Lorentz algebra.

Under the action of the Lorentz group a $D=11$ Dirac spinor is
transformed as
\begin{equation}
\delta\theta=\frac 14 \omega_{\mu\nu}\Gamma^{\mu\nu}\theta.
\end{equation}
Since the $\Gamma^{\mu\nu}$ matrices are real, the reality condition
$\theta^*=\theta$ is compatible with Eq. (A10) which defines a
Majorana spinor. To construct Lorentz-covariant bilinear
combinations, note that
\begin{eqnarray}
&& \delta\bar\theta=-\frac 14\omega_{\mu\nu}
\bar\theta\Gamma^{\mu\nu},\cr
&& \bar\theta\equiv\theta^{\rm T}C.
\end{eqnarray}
Then the combination $\bar\psi\Gamma^\mu\theta$ is a vector under the
action of the $D=11$ Lorentz group
\begin{equation}
\delta(\bar\psi\Gamma^\mu\theta)=-{\omega^\mu}_\nu
(\bar\psi\Gamma^\mu\theta).
\end{equation}
In various calculations the following properties:
\begin{eqnarray}
&& \bar\psi\Gamma^\mu\theta=-\bar\theta\Gamma^\mu\psi, \cr
&& \bar\psi\Gamma^\mu\Gamma^\nu\theta=\bar\theta\Gamma^\nu\Gamma^\mu\psi,\\
&& \bar\psi\Gamma^\mu\Gamma^\nu\Gamma^\rho\theta=
-\bar\theta\Gamma^\rho\Gamma^\nu\Gamma^\mu\psi\nonumber
\end{eqnarray}
are also useful.

It is possible to decompose a $D=11$ Majorana spinor in terms of its
$SO(1,9)$ and $SO(8)$ components. Namely, from Eq. (A8) it
follows that the decomposition
\begin{equation}
\theta=(\bar\theta_\alpha, \theta^\alpha),
\end{equation}
where $\alpha=1,\dots,16$, holds.
Here $\theta$ and $\bar\theta$ are Majorana--Weyl spinors of opposite
chirality with respect to the $SO(1,9)$ subgroup of the $SO(1,10)$ group.
Further, from the third equation in (A8) it follows that in the
decomposition
\begin{equation}
\theta=(\theta_a,\bar\theta'_{\dot a},\theta'_a,\bar\theta_{\dot a}),
\end{equation}
where $a,\dot a=1,\dots,8$, the pairs $\theta_a$, $\theta'_a$ and
$\bar\theta'_{\dot a}$, $\bar\theta_{\dot a}$ are $SO(8)$ spinors of
opposite chirality.

It is convenient to define the $D=11$ light-cone $\Gamma$-matrices
\begin{eqnarray}
&& \Gamma^+=\frac 1{\sqrt 2}(\Gamma^0+\Gamma^9)=
\sqrt 2\left(\mbox{\begin{tabular}{c|c}
0 & $\begin{array}{cc} {\bf 1}_8 & 0\\ 0 & 0\end{array}$\\
\hline
$\begin{array}{cc} 0 & 0\\ 0 & -{\bf 1}_8\end{array}$ & 0
\end{tabular}}\right),\cr
&& \Gamma^-=\displaystyle\frac 1{\sqrt 2}(\Gamma^0-\Gamma^9)=
\sqrt 2\left(\mbox{\begin{tabular}{c|c}
0 & $\begin{array}{cc} 0 & 0\\ 0 & {\bf 1}_8\end{array}$\\
\hline
$\begin{array}{cc} -{\bf 1}_8 & 0\\ 0 & 0\end{array}$ & 0
\end{tabular}}\right),\cr
&& \Gamma^i=\left(\begin{array}{cc} 0 & \Gamma^i\\
\tilde\Gamma^i & 0\end{array}\right), \cr
&& \Gamma^{10}=\left(\begin{array}{cc} {\bf 1}_{16} & 0\\
0 & -{\bf 1}_{16}\end{array}\right),
\end{eqnarray}
where $i=1,\dots,8$. Then the equation $\Gamma^+\theta=0$ has a solution
\begin{equation}
\theta=(\theta_a,0,0,\bar\theta_{\bar a}).
\end{equation}
Besides, under the condition $\Gamma^+\theta=0$ the following
identities:
\begin{eqnarray}
&& \bar\theta\Gamma^+\partial_1\theta=\bar\theta\Gamma^i\partial_1\theta=
\bar\theta\Gamma^{10}\partial_1\theta=0, \cr
&& (\bar\theta\Gamma^\mu\partial_1\theta)\Gamma^\mu\theta=0
\end{eqnarray}
hold.

\end{document}